\documentclass[ACS]{WileyNJD-v1}

\articletype{Article Type}%

\usepackage[utf8x]{inputenc}
 \usepackage[T1]{fontenc}
\usepackage{diagbox}
\usepackage{amsfonts}
\usepackage{amsmath}
\usepackage{amsthm}
\usepackage{indentfirst}
\usepackage{microtype}
\usepackage[squaren]{SIunits}
\usepackage{booktabs}
\usepackage{graphicx}
\usepackage{multirow}
\usepackage{wallpaper}

\usepackage{braket}
\usepackage{caption}
 \makeatletter
 \g@addto@macro\normalsize{%
   \setlength\abovedisplayskip{15pt}
   \setlength\belowdisplayskip{15pt}
   \setlength\abovedisplayshortskip{15pt}
   \setlength\belowdisplayshortskip{15pt}
 }

\makeatother

\begin{document}
\title{Benchmarking vdW-DF first principle predictions against Coupled Electron-Ion Monte Carlo for high pressure liquid hydrogen}

\author[1]{Vitaly Gorelov}

\author[1,2]{Carlo Pierleoni*}

\author[3]{David M. Ceperley}

\authormark{Vitaly Gorelov \textsc{et al}}

\address[1]{\orgdiv{Maison de la Simulation, CEA, CNRS, Univ. Paris-Sud, UVSQ}, \orgname{Universit\'e Paris-Saclay}, \orgaddress{\country{91191 Gif-sur-Yvette, France}}}

\address[2]{\orgdiv{Department of Physical and Chemical Sciences}, \orgname{University of L'Aquila}, \orgaddress{\country{Italy}}}

\address[3]{\orgdiv{Department of Physics}, \orgname{University of Illinois Urbana-Champaign}, \orgaddress{\state{Illinois}, \country{USA}}}

\corres{*Carlo Pierleoni, \email{carlo.pierleoni@aquila.infn.it,carlo.pierleoni@cea.fr}}

%\presentaddress{This is sample for present address text this is sample for present address text}

%\affiliation{
%$^1$Department of Physical and Chemical Sciences, University of L'Aquila, Via Vetoio 10, I-67010 L'Aquila, Italy\\
%$^2$Maison de la Simulation, CEA Saclay, France\\
%$^3$LPMMC, UMR 5493 of CNRS, Universit\'e Grenoble Alpes, F-38042 Grenoble, France\\
%$^4$Institut Laue-Langevin, BP 156, F-38042 Grenoble Cedex 9, France\\
%$^5$Department of Physics, University of Illinois Urbana-Champaign, USA\\ 
%}

%\date{\today}
%\begin{abstract}
\abstract[Summary]{We report first principle results for nuclear structure and optical responses of high pressure liquid hydrogen along two isotherms in the region of molecular dissociation. We employ Density Functional Theory with the vdW-DF approximation (vdW) and we benchmark the results against existing predictions from Coupling Electron-Ion Monte Carlo (CEIMC). At fixed density and temperature, we find that pressure from vdW is higher than pressure from CEIMC by about 10 GPa in the molecular insulating phase and about 20 GPa in the dissociated metallic phase. Molecules are found to be overstabilized using vdW, with a slightly shorter bond length, and with a stronger resistance to compression. As a consequence, pressure dissociation along isotherms using vdW is more progressive than computed with CEIMC. Below the critical point, the liquid-liquid phase transition is observed with both theories in the same density region but the one predicted by vdW has a smaller density discontinuity, i.e. a smaller first order character. The  optical conductivity computed using Kubo-Greenwood is rather similar for the two systems and reflects the slightly more pronounced molecular character of vdW. }
%\end{abstract}

\keywords{high pressure hydrogen, liquid-liquid phase transition, liquid structure, quantum monte carlo methods}

\maketitle

\section{Introduction}
Condensed hydrogen at high pressure is interesting because of the interplay between molecular dissociation and metallization in the liquid phase at high temperatures, and between other phase transitions in the solid phase at lower temperatures\cite{McMahon2012a} . First principle methods are the main theoretical tools for investigation of this system. The proximity of metallization and the dispersive character of the molecular interactions in the insulating phase have proven to be challenging for the well-established Density Functional Theory (DFT) based methods\cite{Morales2010,Morales2013,Morales2013liquid,Clay2014,Azadi2013,Knudson2015}. Predictions from various approximations are in qualitative agreement but there are substantial quantitative differences which makes it difficult to make accurate predictions without benchmarking DFT results against more fundamental theories.  At the same time, the simple electronic structure of hydrogen makes it the ideal candidate for developing new first principle methods based on more fundamental many-body theories such as those based on Quantum Monte Carlo techniques. Two similar methods have appeared in the last decade and were applied to study compressed hydrogen: the Coupled Electron-Ion Monte Carlo (CEIMC) method \cite{Pierleoni2006,Morales2010,Tubman2015,Pierleoni2016} and the Quantum Monte Carlo Molecular Dynamics (QMCMD) \cite{Attaccalite2008,Mazzola2018}. They are both based on solving the electronic problem by ground state Quantum Monte Carlo methods but they differ in the way they sample the nuclear configuration space\footnote{CEIMC sample the nuclear configuration space by Metropolis Monte Carlo which requires computing electronic total energies, while QMCMD implements a Langevin Dynamics based on electronic forces on the nuclei.}.
The advantage of QMC over DFT is in its variational character: very different implementations of QMC, i.e. different forms of the many body wave function, can be directly compared and ranked based on the value of their total energy
\footnote{The variance of the total energy is also an important quantity to compare since it decreases when the quality of the wave function is improved. Therefore a better wave function has a lower energy and a smaller variance.}.
After several attempts, these two methods are now in agreement about the occurrence and the location of the liquid-liquid phase transition (LLPT) in hydrogen \cite{Mazzola2018}.  This reinforces the idea that QMC can be used to assess the accuracy of  DFT functionals for finite temperature predictions. 

Previous detailed investigations of the accuracy of several DFT approximations against ground state QMC for both crystalline and liquid static configurations \cite{Clay2014} found van-der-Waals (vdW-DF) \cite{Dion2004} and Heyd-Scuseria-Ernzerhof (HSE) \cite{Heyd2005} approximations to be the best candidates among the ones considered, with energy biases within fractions of mH/atoms and pressure biases within 10-20 GPa. A CEIMC investigation of the LLPT in hydrogen \cite{Pierleoni2016} confirmed these expectations: first principle molecular dynamics (FPMD) with vdW-DF and HSE approximations predict LLPT transitions in closer agreement with the results from CEIMC than other approximations. 
Differences in the proton-proton pair correlation functions between CEIMC and FPMD are observed \cite{Morales2010,Knudson2015,Pierleoni2017}. These differences could provide insight into the dissociation mechanism. 

In this paper we compare the thermodynamics and local nuclear structure from first principle simulations with vdW-DF and CEIMC, along two isotherms of hydrogen across the molecular dissociation region. These isotherms are above and below the critical temperature of the LLPT. A comparison of the optical conductivity within the Kubo-Greenwood framework is made. We only consider vdW-DF and disregard HSE since the computational requirements of HSE are much larger than of vdW-DF, while the expected accuracy is comparable \cite{Clay2014}.

The paper is organized as follows. First we briefly describe the methods employed in section \ref{sec:method}. Then we report our results for the thermodynamics and for the local nuclear structure in section \ref{sec:results}. In the following section \ref{sec:optics} we compare the optical conductivity and finally we discuss the results and draw our conclusions.

\section{Method}
\label{sec:method}
In this paper we used Metropolis Monte Carlo sampling for the nuclear degrees of freedom and, since the focus is on the accuracy of the potential energy surface, we limit our discussion to classical protons. The electronic energies used in the Metropolis sampling arise either from ground state QMC in the CEIMC \cite{Pierleoni2006,McMahon2012a}, or from ground state Density Functional Theory with the vdW-DF approximation. In both methods the sampling is performed by a force-biased MC algorithm with DFT-forces \cite{Allen1987}. Given a nuclear configuration $\vec{R}$ of the $N_p$ protons (here vectors are in the $3 N_p$ dimensional space), a new configuration $\vec{R}'$ is proposed by a drifted random displacement according to
\begin{equation}
\vec{R}' = \vec{R}+ h \vec{F}(\vec{R}) +\vec{\xi}
\end{equation}
where $\vec{F}(\vec{R})$ is the total force acting on the protons, $h$ is an adjustable parameter with suitable dimensions, and $\vec{\xi}$ a gaussian distributed random vector with the properties
\begin{equation}
\langle \vec{\xi} \rangle=0 \qquad \qquad \langle \vec{\xi}~ \vec{\xi}' \rangle= 2 ~h ~\mathbf{1} ~\delta_{\vec{\xi},\vec{\xi}'}
\end{equation}
where $\mathbf{1}$ is the identity matrix and $\delta_{\vec{\xi},\vec{\xi}'}$ indicates that the random noises at different steps are uncorrelated. Then the sampling probability is
\begin{equation}
G(\vec{R}\rightarrow \vec{R}'; h)\propto e^{-\frac{\left[\vec{R}'-\vec{R}-h\vec{F}(\vec{R})\right]^2}{4h}}
\end{equation}
while the acceptance probability, satisfying detailed balance, is
\begin{equation}
A(\vec{R}\rightarrow \vec{R}')=min\left[1,q(\vec{R}\rightarrow \vec{R}')\right] \qquad\qquad q(\vec{R}\rightarrow \vec{R}')=
\frac{e^{-\beta E(\vec{R}')}G(\vec{R}'\rightarrow \vec{R}; h)}{e^{-\beta E(\vec{R})}G(\vec{R}\rightarrow \vec{R}'; h)}.
\end{equation}
In the last equation $E(\vec{R})$ is the total energy of nuclear configuration $\vec{R}$. 
Within the DFT framework, this scheme can be used straightforwardly, both forces and energies are from DFT-vdW-DF functional. 
This strategy allows one to have a relatively high acceptance probability and also large moves even for system of hundreds of protons, hence an efficient sampling of the nuclear configurational space. 

When solving the electronic problem with QMC, energy and forces have statistical noise that must taken into account to avoid biasing the sampling. In CEIMC the noise over electronic energy is accounted for with the Penalty method\cite{Ceperley1999,Pierleoni2006} while forces are not computed. Our present CEIMC implementation uses a two-level Metropolis scheme in which we first propose a new configuration according to the scheme described above based on DFT energy and forces, if the new configuration $\vec{R}'$ is accepted we perform a second Metropolis acceptance test according to
\begin{equation}
A_2(\vec{R}\rightarrow \vec{R}')=min\left[1,q_2(\vec{R}\rightarrow \vec{R}')\right] \qquad\qquad q_2(\vec{R}\rightarrow \vec{R}')=
\frac{e^{-\beta E(\vec{R})}}{e^{-\beta E(\vec{R}')}}~e^{-\left[\beta\delta(\vec{R}',\vec{R})+\beta^2 u(\vec{R}',\vec{R})\right]}
\end{equation}
where $\delta(\vec{R}',\vec{R})$ indicates the estimate of the difference between the QMC energies of configurations $\vec{R}'$ and $\vec{R}$, and $u(\vec{R'},\vec{R})$ is a positive term, hence a penalty in the acceptance coming from the noise which, to first approximation, can be replaced by the estimate of the variance of the energy difference \cite{Ceperley1999,Pierleoni2006}. 

We do not report new CEIMC calculations but use past calculations as a reference. Details about these calculations are reported in refs. \cite{Pierleoni2016,Pierleoni2017,Rillo2018}. We do report new calculations using DFTMC employing the vdW-DF XC functional and the plane augmented-wave (PAW) method \cite{Blochl1994} with an energy cutoff of 40 Rydberg. In order to have a  direct comparison with the CEIMC results, the DFT calculations have been performed with exactly the same protocol employed in CEIMC. Specifically we did simulations of periodic systems of $N_p=54$ protons, with a 4x4x4 Monkhorst-Pack grid of k-points. This choice ensures that both results are affected by the same finite size error at the single electron level. In CEIMC, many-body size effects are estimated and thermodynamic results corrected accordingly\cite{Pierleoni2016,Holzmann2016}\footnote{Many-body size effects on the local nuclear structure are negligible.}. A further check that residual finite size effects in the DFT calculations are small is provided by direct comparison with calculations for a system of $N_p=128$ protons as detailed in the next section. The DFTMC calculations are performed with our in-house CEIMC code (BOPIMC) which uses the PWSCF\cite{Giannozzi2017} package as a DFT solver. 

The electrical conductivity was computed using the Kubo-Greenwood (KG) \cite{Kubo1957a,Greenwood1958} formula on 40 randomly selected configurations. The configurations are taken from DFTMC and CEIMC runs with $N_P=54$ protons for a given density ($\rho$) and temperature ($T$).  
A post-processing tool in the QuantumEspresso suite\cite{Giannozzi2017}, KGEC \cite{Calderin2017a} was employed. For these calculations we used an 8x8x8 Monkhorst-Pack grid of k-points to get a better convergence in the low-$\omega$ part of the conductivity and included 48 bands to converge in the high-$\omega$ part up to $\omega\simeq 13 eV$. 

\section{Results}
\label{sec:results}
\begin{figure}
\centering
\includegraphics[width=0.9\columnwidth]{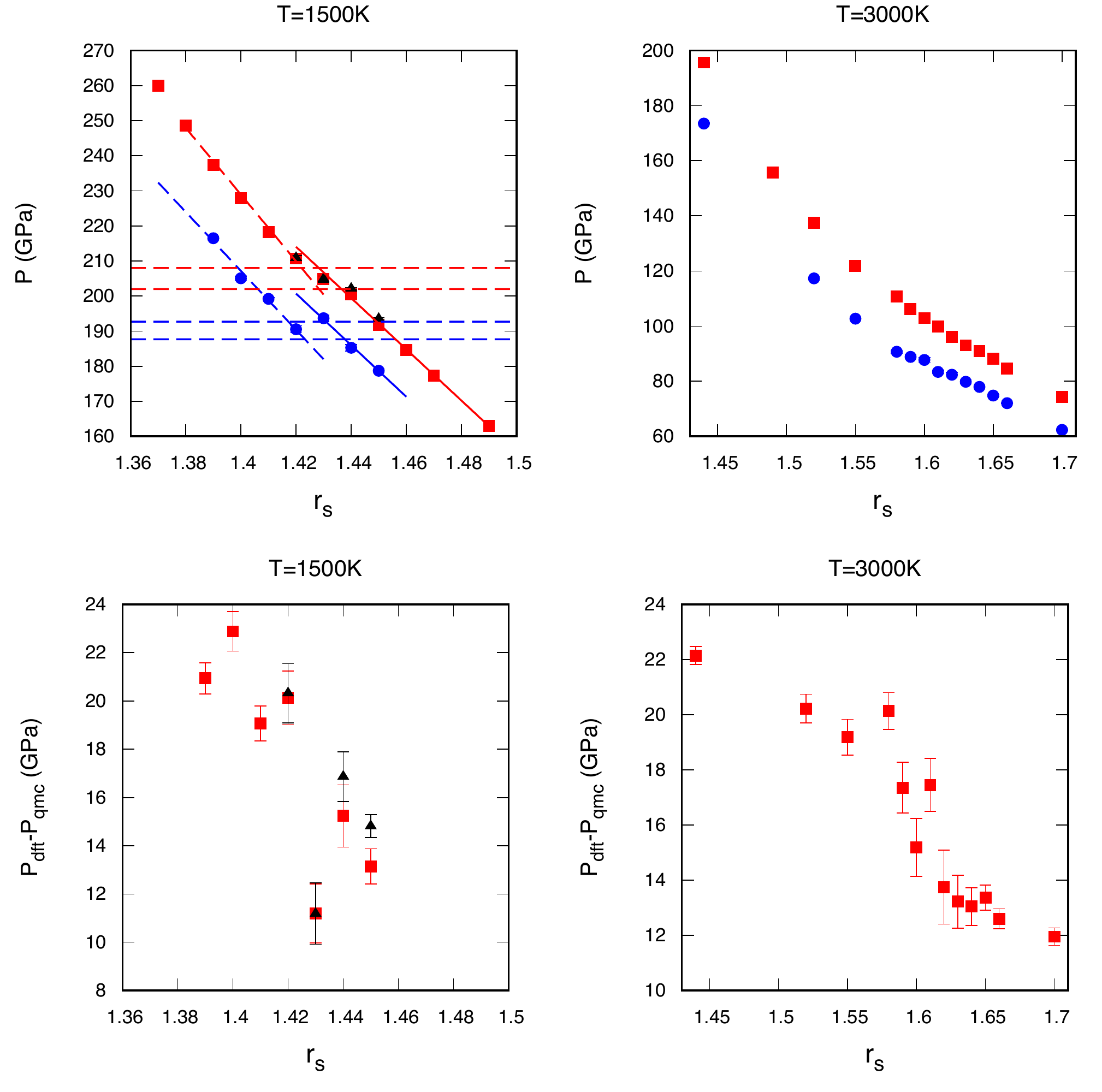}
\caption{EOS along the two isotherms investigated. The upper panels report the pressure versus $r_s$ from CEIMC (blue circles) and from DFTMC with $N_p=54$ (red squares). At $T=1500K$  DFTMC for systems with $N_p=128$ are reported (black triangles). In the lower panels the difference between DFT and QMC pressures is shown. Symbols are as in the upper panels.}
\label{fig:EOS}
\end{figure}    
We performed simulations along two isotherms $T=1500K, 3000K$, one below and one above the critical temperature of the LLPT. (The precise value of T$_c$ is yet to be determined accurately.) In fig. \ref{fig:EOS} we compare the equation of state (EOS) along the two isotherms from CEIMC and from DFTMC-vdW-DF. The data shown in fig \ref{fig:EOS} for CEIMC are from VMC and include finite size corrections and RQMC corrections as explained in the SM of ref. \cite{Pierleoni2016} (see also \cite{Holzmann2016}). The corresponding numerical values are given in the SM of ref. \cite{Pierleoni2016}.
At the lower temperature, we observe a weakly first order phase transition in both theories in the same range of densities. However the width of the density plateau, obtained as the shift between the two branches of the EOS from fitting the data, is about two times larger in CEIMC than in DFT, signaling a stronger first order character of the transition in QMC than in DFT. Residual finite size effects, that could arise from the limited size of our systems, in particular at the phase transition, are negligibly small as shown in the upper left panel of the figure. The pressure from vdW-DF is always larger than from CEIMC. Moreover the accuracy of the vdW-DF predictions for the EOS is not uniform across molecular dissociation. This is shown in the two lower panels of the figure where we report the difference between DFT and QMC pressures. At $T=1500K$, molecular dissociation occurs suddenly between $r_s=1.43$ and $r_s=1.42$ where the pressure difference jumps from $\sim 10-14$GPA to $\sim 20-22$GPa. At $T=3000K$ molecular dissociation, as well as the change in pressure difference, are more gradual with density. However, the observed pressure difference is in the same range of $12GPa \leq \Delta P\leq 22GPa$.  

\begin{figure}
  \centering
  \includegraphics[width=0.8\columnwidth]{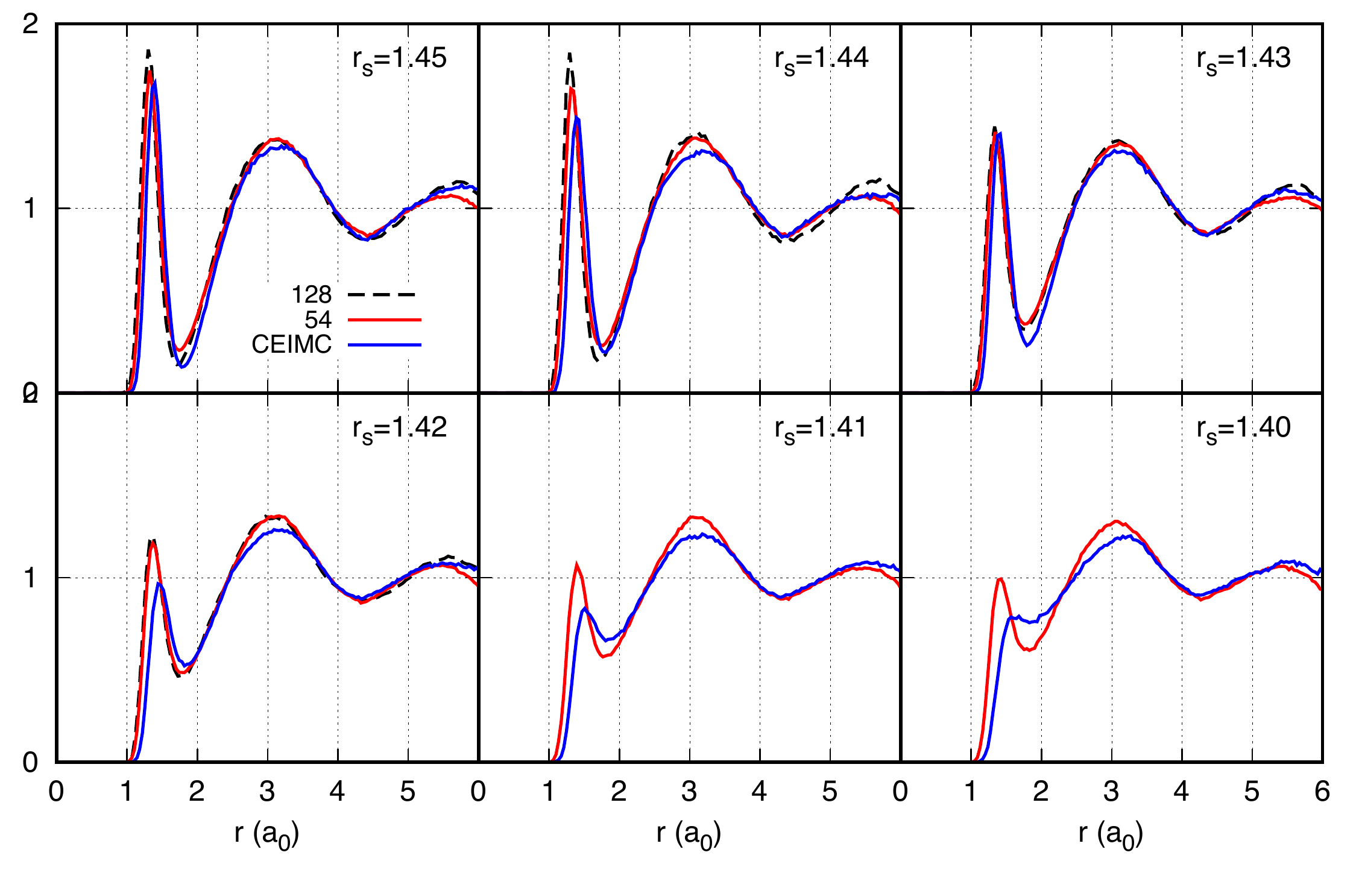}
  \caption{Proton-proton radial distribution functions along the $T=1500K$ isotherm. Comparison between CEIMC (blue lines) and DFTMC-vdW-DF with $N_p=54$ (red lines). At $r_s=1.45, 1.44, 1.43, 1.42$ results for systems of $N_p=128$ with DFTMC-vdW-DF are also reported (black dashed line).}
  \label{fig:gr1500}
\end{figure} 

\begin{figure}
  \centering
  \includegraphics[width=0.8\columnwidth]{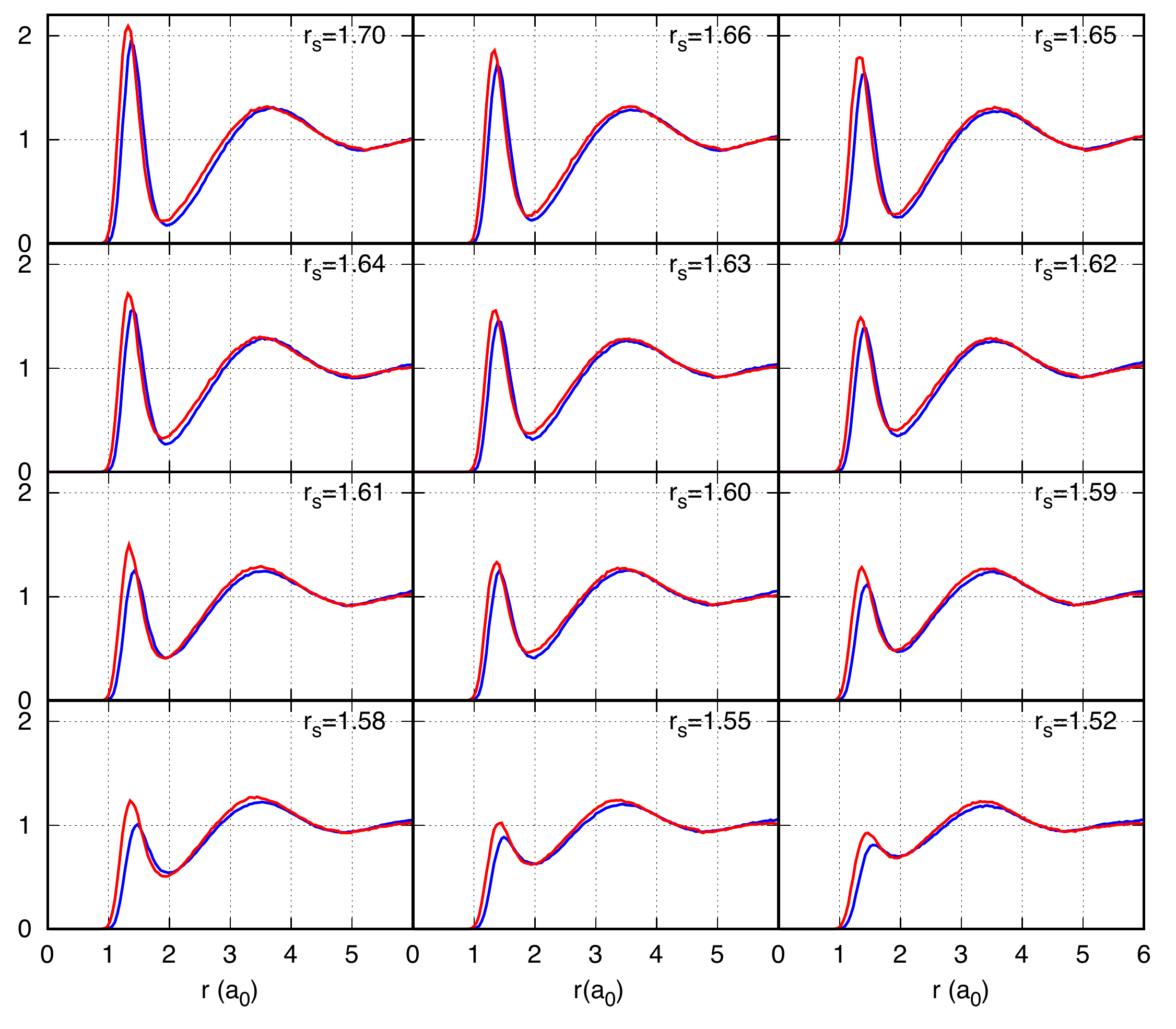}
  \caption{Proton-proton radial distribution function along the $T=3000K$ isotherm. Comparison between CEIMC (blue lines) and DFTMC-vdW-DF with $N_p=54$ (red lines).}
  \label{fig:gr3000}
\end{figure} 

\begin{figure}
  \centering
  \includegraphics[width=0.8\columnwidth]{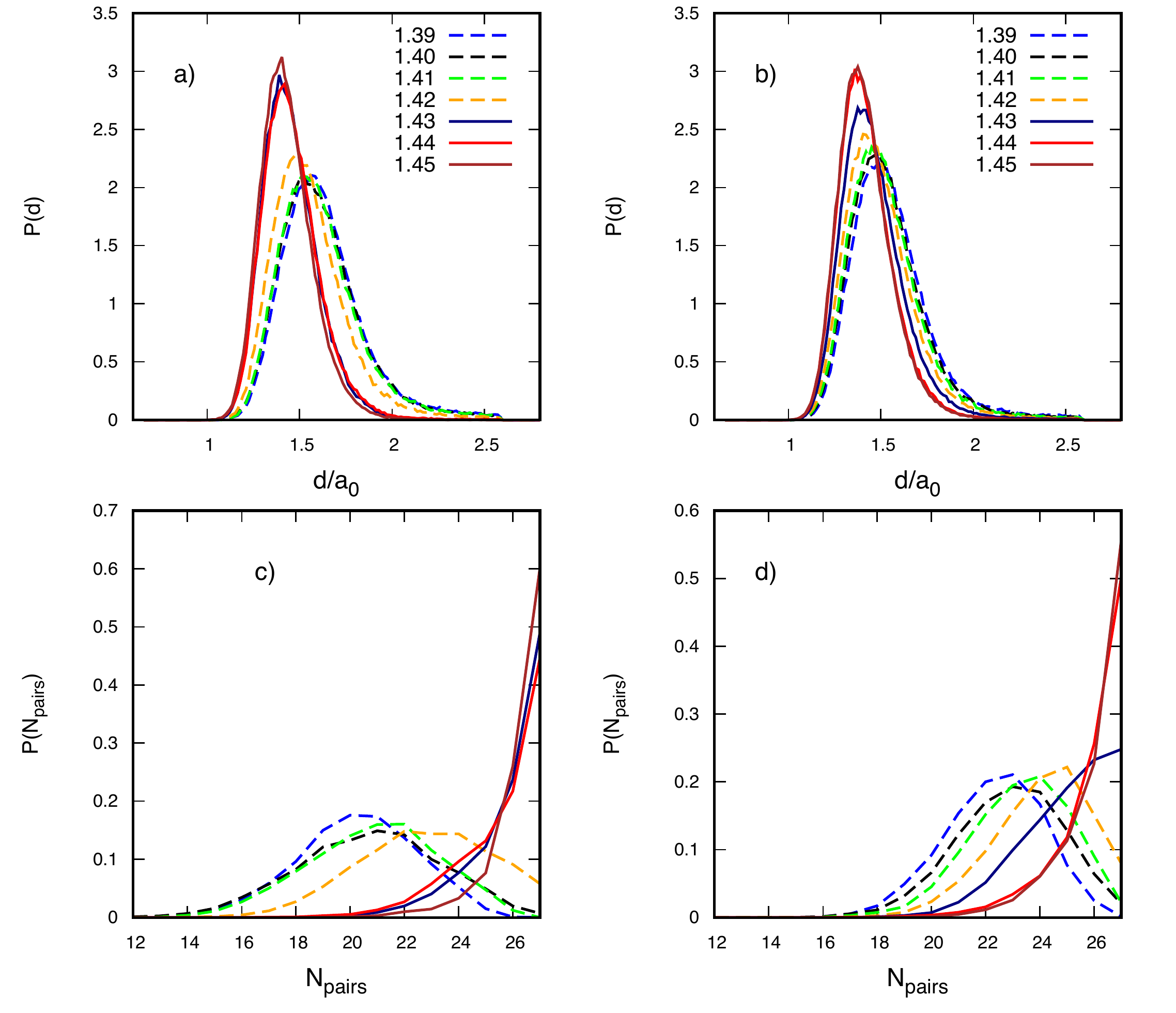}
  \caption{$T=1500K$ isotherm. Bond length distributions obtained with $r_c=2.6$ (panels a) and b)) and the probability of the number of pairs at distance $r\leq 1.8 a_0$ (panels c) and d). CEIMC results panels a) and c); DFTMC results, panels b) and d).}
  \label{fig:pairs1500}
\end{figure} 
\begin{figure}
  \centering
  \includegraphics[width=0.8\columnwidth]{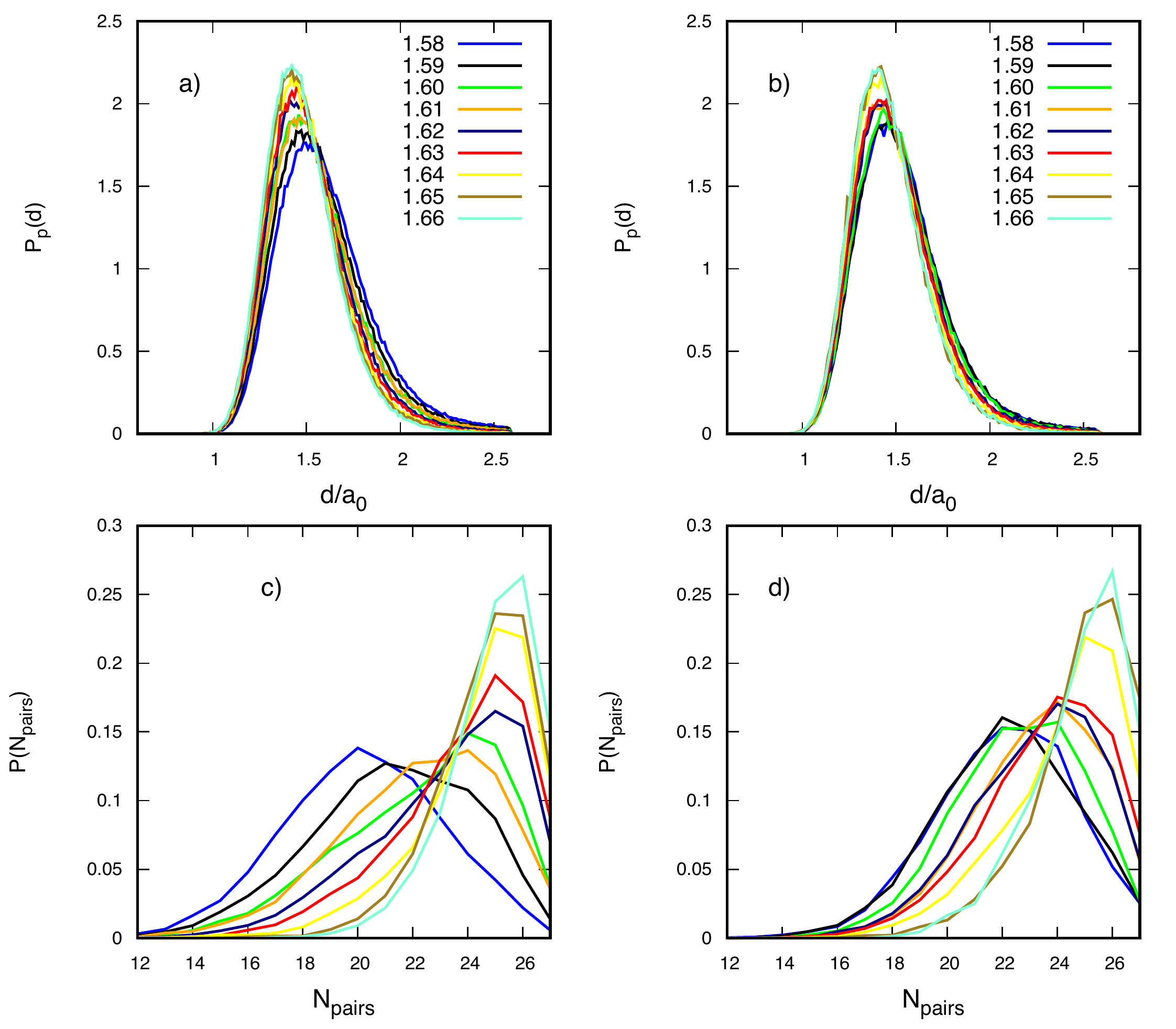}
  \caption{$T=3000K$ isotherm. Bond length distributions obtained with $r_c=2.6$ (panels a) and b)) and the probability of the number of pairs at distance $r\leq 1.8 a_0$ (panels c) and d). CEIMC results panels a) and c), DFTMC results, panels b) and d).}
  \label{fig:pairs3000}
\end{figure} 

In fig. \ref{fig:gr1500} we compare the proton-proton $g(r)$ along the $T=1500K$ isotherm. We observe a qualitative agreement between the two theories, but data from DFTMC exhibit a more pronounced molecular character in the entire density range. The molecular bond length in DFTMC theory is slightly shorter than in CEIMC, and the molecular dissociation is more progressive upon increasing density, in agreement with the smaller density discontinuity observed in the EOS. The same comparison along the $T=3000K$ isotherm is reported in fig. \ref{fig:gr3000}. Here, in both theories, the molecular dissociation is a progressive process. As before, the molecular peak is slightly shorter and the molecular character is slightly more pronounced (more correlation) in DFTMC than in CEIMC. The overall agreement is better than at lower temperature. 
 
In order to investigate in more detail the proton pairing in the system, we have used a cluster algorithm to identify, for any given nuclear configuration, the number of bound pairs. A similar analysis for CEIMC has been already reported in ref. \cite{Pierleoni2017}. Specifically given a proton, the algorithm looks for the closest proton which has not been already paired to another proton at shorter distance. The algorithm find all such pairs within a given cutoff distance. 

In the first analysis we have used a large cutoff distance, $r_c=2.6a_0$ in order to compute the bond length distribution without  artificially cutting off the large distance tail. Results of this analysis for both CEIMC and DFTMC trajectories along the two isotherms are illustrated in the upper panels of figs. \ref{fig:pairs1500} and \ref{fig:pairs3000} where we report the bond distribution function. In each figure, the left panels correspond to the CEIMC results, while the right panels to DFTMC. 
At $T=1500K$ we see a change in behavior for both theories at the LLPT, the discontinuity being more pronounced in CEIMC, but still clearly visible in DFTMC\footnote{Convergence of results at the coexistence density is problematic since the system is observed to switch from one phase to the other several times during the simulation.}. In the molecular phase the observed distribution is rather independent on the specific density, the maximum of probability is at $r=1.40-1.42 a_0$ in CEIMC and at $r=1.36-1.38 a_0$ in DFTMC, and the shape of the distribution is rather similar in the two methods. A similar difference in the molecular bond length between vdW-DF and QMC was previously observed in the crystalline state (see fig. 4 of ref. \cite{Clay2014}). At densities beyond molecular dissociation the bond distribution exhibits a more pronounced density dependence in both theories but remains slightly more localized in DFTMC than in CEIMC (the maximum of probability in DFTMC is at $r\sim 1.49a_0$ while it is at $r\sim 1.58 a_0$ in CEIMC). 
At $T=3000K$ a more progressive change of the bond distribution is observed over the investigated density range. The two theories have similar behavior but again the CEIMC results exhibit a stronger density dependence. 

In the second analysis we used a cutoff distance of $r_c=1.8 a_0$(the first minimum of the $g(r)$) and we computed the distribution in the number of the bonded pairs found within this cutoff. Results for both CEIMC and DFTMC along the two isotherms are illustrated in the lower panels of figs. \ref{fig:pairs1500} and \ref{fig:pairs3000}. Also this distribution exhibits a sudden change of shape at the LLPT along the $T=1500K$ isotherm, as seen in fig. \ref{fig:pairs1500}. Again the two theories are in qualitative agreement. More quantitatively, DFTMC shows a more localized distribution in the ``dissociated'' phase after the LLPT compared to CEIMC, hence a stronger residual associated character. At higher temperature the change of the shape of the distribution with density is gradual in both theories. However, as observed at lower temperature, the CEIMC distributions become wider than the DFTMC distributions for increasing density signaling a less associated character.

%\begin{figure}
%\centering
%\begin{minipage}{.5\textwidth}
%  \centering
%  \includegraphics[width=.95\linewidth]{CondRS145T1500}
%  \captionof{figure}{Real part of the average optical conductivity at $r_s=1.45$ and $T=1500K$ from CEIMC and DFTMC. Statistical errors are represented by the thickness of the lines.}
%  \label{fig:sigma145}
%\end{minipage}%
%\begin{minipage}{.5\textwidth}
%  \centering
%  \includegraphics[width=.95\linewidth]{DOSRS145T1500}
%  \captionof{figure}{Average electronic density of states at $r_s=1.45$ and $T=1500K$ from CEIMC and DFTMC.Statistical errors are represented by the %thickness of the lines.}
%  \label{fig:dos145}
%\end{minipage}
%\end{figure}
\begin{figure}
  \centering
  \includegraphics[width=0.8\columnwidth]{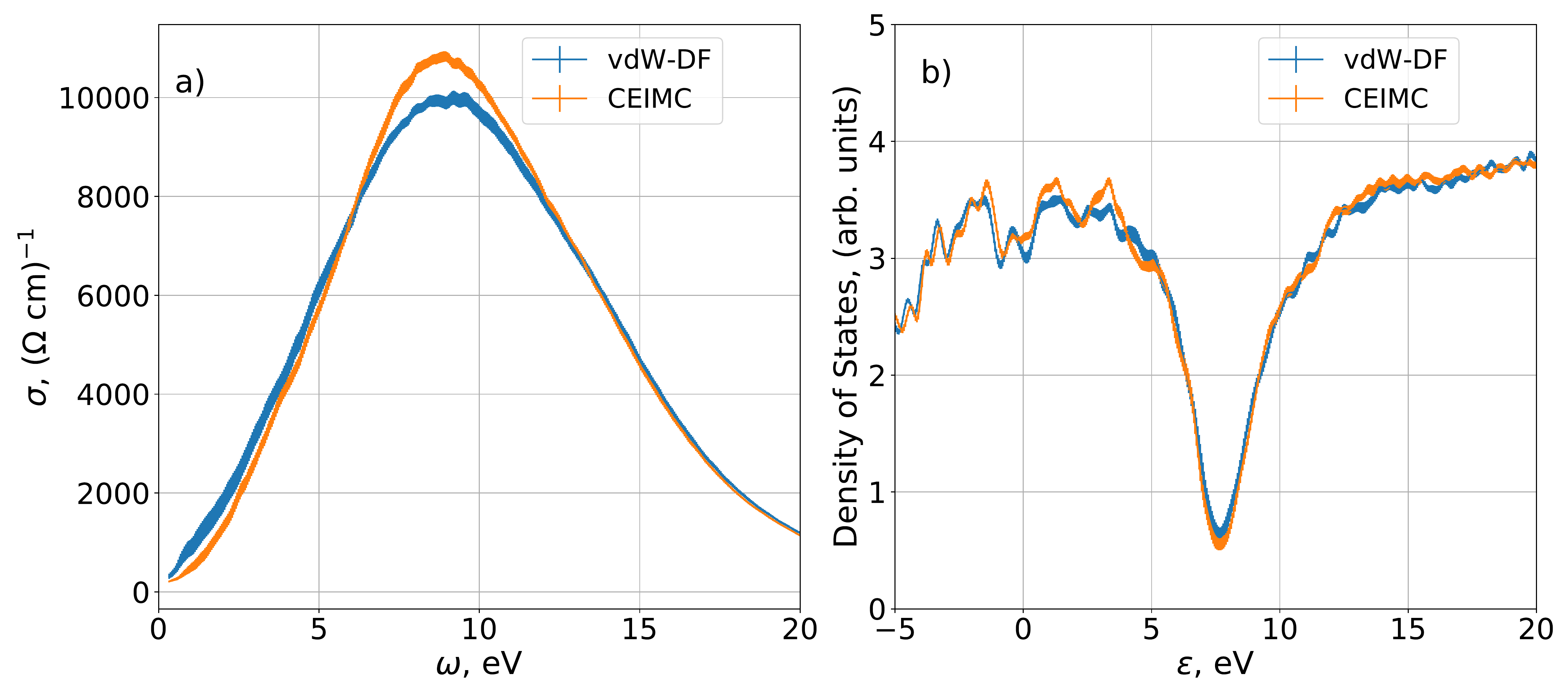}
  \caption{$T=1500K, r_s=1.45$. a) Real part of the average optical conductivity from CEIMC and DFTMC.  b) Average electronic density of states from CEIMC and DFTMC. Statistical errors are represented by the thickness of the lines.}
  \label{fig:sigma145}
\end{figure} 

\section{Optical conductivity}
\label{sec:optics}
Optical properties are relevant since they are the main source of experimental information at these high pressure-high temperature states of liquid hydrogen. In particular recent experiments at NIF \cite{Celliers2018} provided support for the LLPT lines predicted by CEIMC and FPMD-vdW-DF for deuterium on the basis of the optical response of the system\footnote{The LLPT lines from CEIMC and from FPMD-vdW-DF are closer than the experimental resolution.}. A subsequent investigation, within the Kubo-Greenwood framework, of optical properties across molecular dissociation support experimental findings \cite{Rillo2018b}, although other experiments suggest different scenarios \cite{Knudson2015} and the debate is still ongoing.  For the purposes of the present benchmark, we believe it is interesting to check the sensitivity of optical properties to the details of the local liquid structure. To this aim we have computed the optical conductivity at three distinct phase points, $(r_s=1.45, T=1500K)$, $(r_s=1.40, T=1500K)$ and $(r_s=1.52, T=3000K)$ for liquid structure from both methods. At each state point and for each simulation method, we have considered 40 independent nuclear configurations generated during the simulation run, and for each nuclear configuration we have computed the optical conductivity with the Kubo-Greenwood single electron formalism, assuming Kohn-Sham eigenstates from the vdW-DF XC functional. Details are given in section \ref{sec:method}. Statistical averages and errors are obtained over the sample of configurations. Figure \ref{fig:sigma145}a) compares the real part of the average optical conductivity from CEIMC and from DFTMC-vdW-DF at the first state point, while fig. \ref{fig:sigma145}b) reports the average electronic density of states from both theories. Results from the two methods are in good agreement showing, once more, the similarity between the generated nuclear structures. However, we observe a slightly larger density of states at the Fermi level, which correlates with a slightly higher conductivity at small energies, from DFTMC-vdW-DF than from CEIMC. Conversely around the maximum of conductivity at $\omega\simeq 9 eV$, CEIMC data are slightly higher than DFTMC-vdW-DF data. %\CP{We think this is because of a sum rule, $\int=1$, that holds for insulating states}.  
Note that this state point is at the edge of the metallic regimes: the energy gap is closed but the density of states at the Fermi level is rather small, resulting in a very small DC conductivity.

\begin{figure}
  \centering
  \includegraphics[width=0.8\columnwidth]{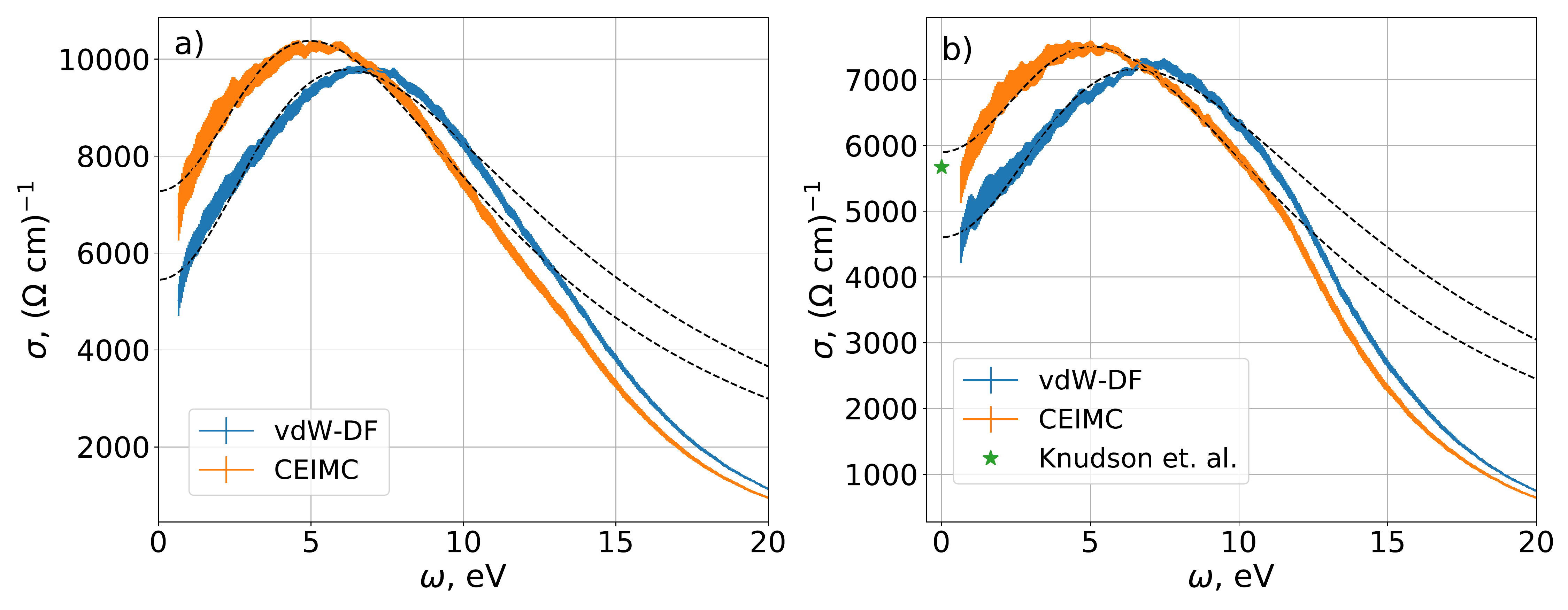}
  \caption{a) Real part of the average optical conductivity at $r_s=1.40$ and $T=1500K$ from CEIMC and DFTMC.  b) Real part of the average optical conductivity at $r_s=1.52$ and $T=3000K$ from CEIMC and DFTMC.  Dashed lines represent a Drude-Smith fit to data. Statistical errors are represented by the thickness of the lines.}
  \label{fig:sigma140-152}
\end{figure}

A similar analysis into the conducting regime is presented in figs. \ref{fig:sigma140-152} where optical conductivities for the other two investigated state points are reported. We observe similar comparisons at both state points: CEIMC conductivity is larger at small energies and the maximum is at slightly lower energy than the DFTMC-vdW-DF one. This difference can be qualitatively justified as arising from the less ``molecular'' character of the CEIMC configurations (see figs. \ref{fig:gr1500} and \ref{fig:gr3000}), hence from a weaker electronic localization. Analysis of the conductivities in terms of the Drude-Smith model\cite{Smith2001,Cocker2017a} provides $\sigma_{dft}(\omega=0)=5450(50), \sigma_{qmc}(\omega=0)=7280(70) (\Omega cm)^{-1}$ at $T=1500 K$ $r_s=1.40$ and $\sigma_{dft}(\omega=0)=4600(40), \sigma_{qmc}(\omega=0)=5900(40) (\Omega cm)^{-1}$ at $T=3000 K$ $r_s=1.52$ for vdW-DF and CEIMC respectively. In figure \ref{fig:sigma140-152}b) we also report a recent result for the DC conductivity from FPMD with vdW-DF XC approximation\cite{Knudson2018} at a state point close to our present one\footnote{The comparison is shown to support our present analysis. We believe that the value is slightly larger than our own because the density of that state point is slightly larger.}.

\section{Conclusions}
\label{sec:conclusions}
In conclusion, we have presented a detailed comparison between CEIMC and DFTMC with vdW-DF approximation for high pressure liquid hydrogen across the interesting region where pressure induced molecular dissociation occurs, either as a continuous process at high temperature, or through a first order phase transition at lower temperature. We have computed the EOS, the local nuclear structure of the fluid and the optical conductivity at several state points. We observe that pressure from vdW-DF is higher than from CEIMC by about $10GPa$ in the molecular phase and about $20GPa$ in the dissociated phase, indicating a loss of accuracy of the approximation upon molecular dissociation. Inspection of the local nuclear structure, as monitored by the proton-proton $g(r)$, shows that the molecular character of the fluid is more pronounced with DFTMC-vdW-DF and molecules have more resistance to compression and are more difficult to dissociate.
Despite these quantitive difference, our investigation demonstrate once more the good quality of vdW-DF functional to study high-pressure hydrogen in this interesting region of its phase diagram.

%\vskip 2cm
\section*{Acknowledgments}
V. G. and C.P. were supported by the Agence Nationale de la Recherche (ANR) France, under the program ``Accueil de Chercheurs de Haut Niveau 2015'' project: HyLightExtreme. D.M.C. was supported by DOE Grant NA DE-NA0002911 and by the Fondation NanoSciences (Grenoble).  Computer time was provided by PRACE Project 2016143296 and by an allocation of the Blue Waters sustained petascale computing project, supported by the National Science Foundation (Award OCI 07- 25070) and the State of Illinois, and by the HPC resources from GENCI-CINES under the allocation 2018-A0030910282.

\bibliography{dottorato}

\end{document}